\documentclass{emulateapj}

\newcommand{\ionCaT}{\ion{Ca}{2}}

\shorttitle{Towards a Solution for the Calcium Puzzle}
\shortauthors{Michielsen et al.}

\begin{document}

\title{Towards a Solution for the Ca\,II Triplet Puzzle~:~Results
  from Dwarf Elliptical Galaxies}

\author{Dolf Michielsen} \affil{School of Physics and Astronomy, University of
  Nottingham, University Park, NG7~2RD Nottingham, UK}

\author{Mina Koleva\altaffilmark{1} and Philippe Prugniel\altaffilmark{2}}
\affil{Universit\'e de Lyon, Lyon, France ; Universit\'e Lyon~1, Villeurbanne,
  F-69622, France ; Centre de Recherche Astronomique de Lyon, Observatoire de
  Lyon, F-69561, France ; CNRS, UMR 5574 ; Ecole Normale Sup\'erieure de Lyon,
  Lyon, France}

\author{Werner~W. Zeilinger} \affil{Institut f\"{u}r Astronomie,
  Universit\"{a}t Wien, T\"{u}rkenschanzstrasse 17, A-1180 Wien, Austria}

\author{Sven De Rijcke\altaffilmark{3} and Herwig Dejonghe}
\affil{Sterrenkundig Observatorium, Ghent University, Krijgslaan 281, S9,
  B-9000 Ghent, Belgium}

\author{Anna Pasquali} \affil{Max-Planck-Institut f\"ur Astronomie,
  Koenigstuhl 17, D-69117 Heidelberg, Germany}

\author{Ignacio Ferreras} \affil{Physics Department, King's College London,
  Strand, London WC2R~2LS, UK}

\and

\author{Victor~P. Debattista\altaffilmark{4}} \affil{Centre for Astrophysics,
  University of Central Lancashire, Preston PR1~2HE, UK}

\altaffiltext{1}{Department of Astronomy, St.~Kl.~Ohridski University of
  Sofia, BG-1164 Sofia, Bulgaria}

\altaffiltext{2}{GEPI Observatoire de Paris-Meudon, 5 place Jules Janssen,
  Meudon, F-92195, France}

\altaffiltext{3}{Research Postdoctoral Fellow of the Fund for Scientific
  Research, Flanders, Belgium (FWO).}

\altaffiltext{4}{RCUK Academic Fellow}

\begin{abstract}
  We present new estimates of ages and metallicities, based on
  FORS/VLT optical (4400--5500\,\AA) spectroscopy, of 16 dwarf
  elliptical galaxies (dE's) in the Fornax Cluster and in Southern
  Groups. These dE's are more metal-rich and younger than previous
  estimates based on narrow-band photometry and low-resolution
  spectroscopy. For our sample we find a mean metallicity ${\rm [Z/H]}
  = -0.33$\,dex and mean age 3.5\,Gyr, consistent with similar samples
  of dE's in other environments (Local Group, Virgo). Three dE's in
  our sample show emission lines and very young ages. This suggests
  that some dE's formed stars until a very recent epoch and were self-enriched
  by a long star formation history. Previous observations of large
  near-infrared ($\sim 8500$\,\AA) \ionCaT\ absorption strengths in
  these dE's are in good agreement with the new metallicity estimates,
  solving part of the so-called Calcium puzzle.
\end{abstract}


\section{Introduction}

The observed strength of the near-infrared \ionCaT\ triplet absorption
lines in early-type galaxies has presented astronomers with an
interesting puzzle over the past couple of years. \citet{cenarro01}
defined a new CaT* index, carefully correcting for the underlying H
Paschen absorption. Whereas other metallicity tracers, such as Mg$_2$,
correlate with velocity dispersion $\sigma$, it was found that CaT*
\emph{anti-correlates} with $\sigma$ in elliptical galaxies (E's) and
in bulges of spiral galaxies \citep{saglia02, cenarro03,
falcon03}. Population synthesis model predictions also show that, for
sub-solar metallicities, CaT* should be sensitive to metallicity but
virtually independent of age, while at super-solar metallicity, the
CaT* saturates \citep{vazdekis03}. However, taking metallicities
estimated from optical spectra, \citet{saglia02} reported that the
measured CaT* values in E's are smaller by 0.5\,{\AA} than those
predicted by population synthesis models. \citet{michielsen03}
(hereafter Paper~1) showed that the anti-correlation of CaT* with
$\sigma$ extends into the dwarf elliptical (dE) regime.  These dE's
were expected to have metallicities of the order of [Z/H]$\,\sim -1$,
and ages of the order of 10\,Gyr \citep{heldmould94, rakos01}. The
measured CaT* values were significantly larger than those expected for
such old, metal-poor stellar systems. 

All of the proposed solutions to this conundrum, such as variations of
the initial-mass function or the calcium yield as a function of
metallicity or velocity dispersion, require considerable fine-tuning.
None of them satisfactorily explains both the small CaT* in E's and
the large CaT* in dE's without creating other difficulties, e.g.\ with
the FeH~$\lambda$9916 index values observed in bright ellipticals (FeH
is strong in dwarf stars but nearly absent in giants) and with stellar
mass-to-light ratios \citep{saglia02, cenarro03}. However, the stellar
populations of E's and dE's are most likely not single-age,
single-metallicity populations, or SSPs, as was implicity assumed in
essentially all age and metallicity estimates \citep[see
e.g.][]{pasquali05}. Moreover, the stars that dominate the blue
spectral range (mostly hot dwarf stars) do not necessarily have the
same mean ages/metallicities as the stars producing the red light
(mostly cool giants). These issues, together with systematic
uncertainties inherent to population synthesis tools, potentially
contribute to the CaT puzzle.

Because of their low surface brightness, accurate estimates of the
ages and metallicities of dE's are still scarce. Recent studies of
dE's in the Virgo cluster \citep{g03, v04} report younger ages and
higher metallicities than found in Fornax dE's. In Paper~1, the ages
and metallicities of the dE's were taken from the literature, where
low-resolution, modest S/N spectroscopic \citep{heldmould94} or
narrow-band photometric \citep{rakos01} techniques were used. As a
sanity check, we have now acquired high-resolution, high S/N optical
spectra with FORS/VLT of all the dE's for which we presented CaT*
measurements in Paper~1.  This puts us in a position where we can for
the first time compare the CaT* measurements with model predictions
based on robust age and metallicity estimates.

\section{Observations and data reduction} \label{obs}

The sample consists of 7 dE's in the Southern NGC\,5044, NGC\,5898 and Antlia
Groups, and 9 dE's in the Fornax Cluster. All the observations were carried
out in Service Mode at the ESO-VLT, in seeing conditions between $0.6
-0.9$\arcsec~FWHM. Integration times varied between 2 and 7 hours per galaxy.
The Group dE's were observed in April and May 2005 using FORS2 with
\texttt{GRISM\_600B} and a slit width of 0.5\arcsec.  This results in a
wavelength range of $3300 - 6200$\,\AA\ at a 3.0\,\AA\ (FWHM) resolution.  The
Fornax dE's were observed in December 2005 and January 2006 using FORS1 with
the holographic \texttt{GRISM\_1200g} and a slit width of 1.0\arcsec.  This
results in a wavelength range of $4350 - 5530$\,\AA\ at a 2.6\,\AA\ (FWHM)
resolution. The data reduction was carried out using the ESO-MIDAS
package\footnote{The image processing package ESO-MIDAS is developed and
  maintained by the European Southern Observatory.}. The spectra for each
galaxy were bias-subtracted, flat-fielded, corrected for cosmic-ray events,
wavelength calibrated and co-added. Then the sky was subtracted and the
spectra were extinction-corrected and flux-calibrated using spectrophotometric
standard stars. For the purpose of this Letter, we extracted 1D spectra over an
aperture of radius $R_e/8$ \citep[or 1\arcsec\ for galaxies with $R_e <
8$\arcsec;][]{derijcke05}, to be consistent with the region in which the CaT*
index was measured in Paper~1. These spectra have a S/N $> 75$\,\AA$^{-1}$ in
the wavelength region 4700$-$5500\,\AA. A full description of the data
analysis will be presented in a forthcoming paper (Koleva et al., in
preparation).

\section{Results}  \label{res}

\subsection{Ages and metallicities} \label{agemet}

\begin{figure}
\vspace{5cm}
\special{hscale=75 vscale=75 hsize=570 vsize=150 
         hoffset=-58 voffset=-50 angle=0 psfile="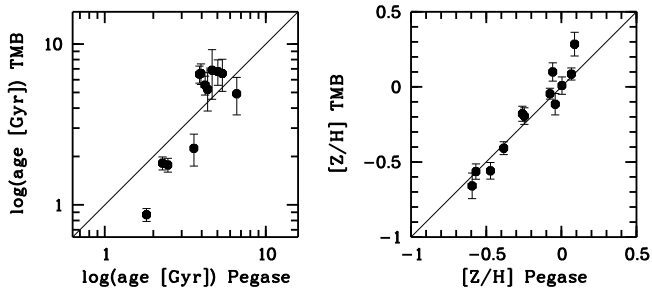"}
    \caption{
      Comparison of SSP-equivalent age and metallicity estimates using
      Lick/IDS indices in combination with TMB03 models and a full spectrum
      fit with Pegase-HR models. The error bars are 1$\sigma$ errors. Both
      approaches clearly agree.\label{fig_compare_agemet}}
\end{figure}

To measure the ages and metallicities of the dE's, we compared our
observations to single-age, single-metallicity population (SSP)
synthesis models. We used two different techniques and models:
\emph{(i)} \citet[TMB03]{TMB03} models in the Lick/IDS index system,
and \emph{(ii)} Pegase-HR models \citep{leborgne04} with the
ELODIE.3.1 stellar library \citep{prugniel07} to which we perform a
full spectrum fit \citep{koleva07}.  To transform our spectra to the
Lick/IDS system, we smoothed our spectra to the Lick/IDS resolution
($\sim 8.4$\,\AA\ FWHM) and measured the H$\beta$, Mg$b$, Fe5270, and
Fe5335 indices \citep{wortheyetal94}. In each setup we have 3 stars in
common with the original Lick/IDS library. This number is too small to
determine systematic offsets, so we applied the offsets to the Jones
library \citep{wo97}. These corrected indices agree within the error
bars with those given in \citet{wo97}. We then compared the indices of
the dE's with TMB03 models with varying [$\alpha$/Fe] abundance
ratios, applying a quadratic interpolation over the 9 nearest SSP
model grid points. The 1$\sigma$ error contours on the derived
quantities are calculated by Monte Carlo simulations taking into
account the measured errors in the indices \citep[see][for a detailed
description]{cardieletal03}. Note that systematic errors introduced by
the conversion to the Lick/IDS system are not taken into account. This
approach yields an estimate for the ages, metallicities and abundance
ratios of the targeted dE's. We find that the dE's have solar
abundance ratios and metallicities in the range $-0.7 < {\rm [Z/H]} <
+0.2$, which justifies the use of the ELODIE.3.1 library to perform
the full-spectrum fit. This yields a second, independent, age and
metallicity estimate for the program dE's. The 1$\sigma$ errors are
computed in a similar way as with the indices. As is obvious from
Figure~\ref{fig_compare_agemet}, both techniques agree, with an rms
difference of 1.63\,Gyr in age and 0.09\,dex in [Z/H]. For a
discussion of possible systematic differences between the model
predictions, we refer interested readers to \citet{leborgne04}. This
makes us confident that our age and metallicity estimates are
robust. The advantage of the full spectrum fitting method is that we
do not have to convert the data to the (low-resolution) Lick/IDS
system and that we can carefully take into account filling by emission
lines.  Emission is present in three dE's in our sample and we did not
study them using the index method. Very low-level [O\,\textsc{III}]
(5007\,\AA) emission was detected in FCC\,150, that would not have
been detected if the resolution or the S/N of the observations were
only slightly worse. The first 5 columns of Table~\ref{tab_agemet}
give, for our sample of dE's, the ages, metallicities and
[$\alpha$/Fe] ratio derived using TMB03, and the age and metallicity
derived using Pegase-HR. In the following we will use Pegase-HR
results. Because the full fit uses all available information in the
spectrum, the errors are typically a factor 2-3 smaller than with
inversions of bi-index grids.

\begin{table*}
 \scriptsize
\begin{center}
 \caption{Ages, metallicities and [$\alpha$/Fe] ratios\label{tab_agemet}}
 \begin{tabular}{l*{7}{r@{\,$\pm$\,}l}}
  \tableline
  \tableline
      & \multicolumn{6}{c}{Lick/IDS Indices - TMB03} & \multicolumn{4}{c}{Spectrum Fit - Pegase-HR} & \multicolumn{2}{c}{HM94} & \multicolumn{2}{c}{R01}\\
Galaxy Name& \multicolumn{2}{c}{Age} & \multicolumn{2}{c}{[Z/H]} & \multicolumn{2}{c}{[$\alpha$/Fe]} & \multicolumn{2}{c}{Age} & \multicolumn{2}{c}{[Z/H]} & \multicolumn{2}{c}{[Fe/H]} & \multicolumn{2}{c}{[Fe/H]} \\
           & \multicolumn{2}{c}{(Gyr)}& \multicolumn{2}{c}{(dex)} & \multicolumn{2}{c}{(dex)}  & \multicolumn{2}{c}{(Gyr)}& \multicolumn{2}{c}{(dex)} & \multicolumn{2}{c}{(dex)} & \multicolumn{2}{c}{(dex)} \\
  \tableline
  FCC\,043 & 1.82 & 0.17 & $-$0.05 & 0.04 &    0.05 & 0.03                               & 2.28 & 0.04 & $-$0.08 & 0.01 & \multicolumn{2}{c}{\dots} & \multicolumn{2}{c}{\dots} \\
  FCC\,046 & \multicolumn{2}{c}{\dots} & \multicolumn{2}{c}{\dots} & \multicolumn{2}{c}{\dots} & 1.15 & 0.02 & $-$1.07 & 0.02 & \multicolumn{2}{c}{\dots} & \multicolumn{2}{c}{\dots} \\
  FCC\,136 & 6.56 & 1.48 & $-$0.18 & 0.05 &    0.08 & 0.05                               & 5.37 & 0.23 & $-$0.26 & 0.01 & \multicolumn{2}{c}{\dots} & $-$0.41 & 0.12 \\
  FCC\,150 & 6.50 & 0.80 & $-$0.41 & 0.04 &    0.00 & 0.05                               & 3.87 & 0.12 & $-$0.38 & 0.01 & $-$0.89 & 0.03          & $-$0.75 & 0.11 \\
  FCC\,204 & 1.77 & 0.17 &    0.09 & 0.04 & $-$0.06 & 0.03                               & 2.46 & 0.05 &    0.07 & 0.01 & \multicolumn{2}{c}{\dots} & \multicolumn{2}{c}{\dots} \\
  FCC\,207 & \multicolumn{2}{c}{\dots} & \multicolumn{2}{c}{\dots} & \multicolumn{2}{c}{\dots} & 1.50 & 0.05 & $-$0.72 & 0.03 & $-$1.19 & 0.05          & $-$1.51 & 0.09 \\
  FCC\,245 & 6.75 & 1.23 & $-$0.66 & 0.09 &    0.13 & 0.08                               & 5.04 & 0.34 & $-$0.59 & 0.03 & $-$1.00 & 0.07          & $-$1.17 & 0.10 \\
  FCC\,266 & 6.56 & 0.97 & $-$0.56 & 0.06 & $-$0.01 & 0.05                               & 3.95 & 0.17 & $-$0.47 & 0.01 & $-$0.84 & 0.07          & $-$0.83 & 0.11 \\
  FCC\,288 & 5.56 & 0.76 & $-$0.56 & 0.05 & $-$0.04 & 0.06                               & 4.20 & 0.24 & $-$0.57 & 0.02 & \multicolumn{2}{c}{\dots} & \multicolumn{2}{c}{\dots} \\
  DW\,1    & 5.23 & 1.42 & $-$0.19 & 0.06 &    0.03 & 0.06                               & 4.35 & 0.21 & $-$0.26 & 0.02 & \multicolumn{2}{c}{\dots} & \multicolumn{2}{c}{\dots} \\
  DW\,2    & \multicolumn{2}{c}{\dots} & \multicolumn{2}{c}{\dots} & \multicolumn{2}{c}{\dots} & 1.06 & 0.02 & $-$0.56 & 0.03 & \multicolumn{2}{c}{\dots} & \multicolumn{2}{c}{\dots} \\
  FS\,29   & 2.25 & 0.51 &    0.10 & 0.06 & $-$0.09 & 0.03                               & 3.57 & 0.15 & $-$0.06 & 0.01 & \multicolumn{2}{c}{\dots} & \multicolumn{2}{c}{\dots} \\
  FS\,75   & 2.45 & 0.58 &    0.16 & 0.07 &    0.02 & 0.03                               & 4.01 & 0.17 &    0.04 & 0.01 & \multicolumn{2}{c}{\dots} & \multicolumn{2}{c}{\dots} \\
  FS\,76   & 4.91 & 1.29 &    0.29 & 0.08 &    0.03 & 0.03                               & 6.59 & 0.19 &    0.09 & 0.01 & \multicolumn{2}{c}{\dots} & \multicolumn{2}{c}{\dots} \\
  FS\,131  & 6.88 & 2.35 & $-$0.12 & 0.07 &    0.04 & 0.06                               & 4.64 & 0.33 & $-$0.04 & 0.02 & \multicolumn{2}{c}{\dots} & \multicolumn{2}{c}{\dots} \\
  FS\,373  & 0.87 & 0.08 &    0.01 & 0.06 &    0.01 & 0.03                               & 1.82 & 0.04 &    0.00 & 0.01 & \multicolumn{2}{c}{\dots} & \multicolumn{2}{c}{\dots} \\
  \tableline
 \end{tabular}
\end{center}
\end{table*}

\subsection{Comparison with literature data}

There is clearly a large disparity between the new age and metallicity
estimates presented in this Letter and the existing estimates from the
literature that needs to be explained. \citet[hereafter HM94]{heldmould94} and
\citet[hereafter R01]{rakos01} studied Fornax dE's and have respectively 5 and
6 galaxies in common with our sample. Their metallicity estimates are listed
in the last two columns of Table~\ref{tab_agemet}. First of all, one of the
problems that HM94 faced at the time was the lack of age and metallicity
dependent population synthesis models. They assumed that the ages of the dE's
are comparable to those of Galactic globular clusters in order to justify the
use of a metallicity scale calibrated on globular clusters. However, most
model bi-index grids are not perpendicular in age and metallicity. At a given
measured metal index, a younger age generally implies a higher metallicity.
HM94 were aware of this shortcoming at the time and cautioned that their
metallicities were probably underestimates. One of the galaxies we have in
common with HM94, FCC\,207, shows emission lines, indicative of on-going star
formation. We find a very young SSP-equivalent age of 1.5\,Gyr. From H$\alpha$
imaging, we know that the emission in FCC\,207 is concentrated in the central
1\arcsec\ \citep{derijcke03}. HM94 did not report emission in FCC\,207 (see
their Figure~1), probably because their technique of nodding the telescope
perpendicular to the slit to sample the whole nuclear region diluted the
emission lines.  They did note that the H$\delta$ absorption of FCC\,207 was
considerably stronger than the (emission-filled) H$\gamma$ and H$\beta$
absorption, but attributed this to measurement errors on H$\delta$.

Secondly, R01 calibrated their narrow-band photometry metallicity scale on the
observations of HM94, re-observing all the HM94 dE's.  In fact, R01 noted that
the metallicity derived from the narrow-band colours underestimates the true
metallicity, and that the off-set is larger if the population is younger.
However they only show 10 and 13\,Gyr models, using the latter to compute
corrections for their metallicity scale. Using corrections derived for
younger ages may reconcile their results with ours.

\subsection{CaT measurements and predictions} \label{cat}

\begin{figure}
\vspace{9cm}
\special{hscale=65 vscale=65 hsize=570 vsize=250 
         hoffset=-35 voffset=-30 angle=0 psfile="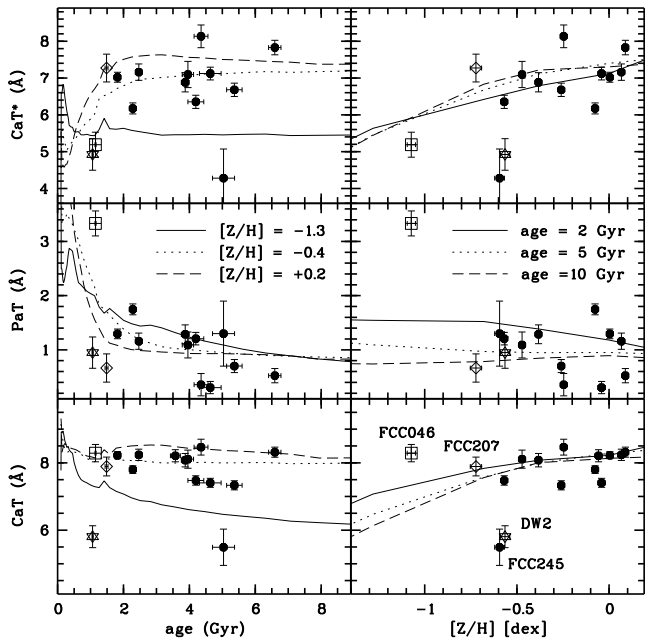"}
    \caption{ CaT, PaT and CaT* versus age and metallicity for the
      Fornax cluster and Southern group dE's. Overlaid are
      \citet{vazdekis03} SSP models with Salpeter IMF. FCC\,046
      (square), FCC\,207 (diamond) and DW2 (star) are the three dE's
      with emission lines. FCC\,245 is the most significant
      outlier.\label{fig_models}}
\end{figure}

\citet{cenarro01} defined a set of new indices that quantify the
strengths of \emph{(i)} the \ionCaT\ lines (CaT), \emph{(ii)} the
Paschen P12, P14, and P17 lines (PaT), and \emph{(iii)} the \ionCaT\
corrected for the absorption contributed by the Paschen P13, P15, and
P16 lines (${\rm CaT*} = {\rm CaT} - {\rm 0.93} \times {\rm PaT}$). In
Figure~\ref{fig_models}, we use our new age and metallicity estimates
to compare the observed CaT, PaT and CaT* strengths to the SSP
predictions of \citet{vazdekis03} with Salpeter initial mass function
(IMF). FS\,75 was not observed in the \ionCaT. For most of the dE's,
the values are now consistent with the model predictions. FCC\,046 has
a high CaT for its age and metallicity. However, this galaxy has
emission lines and shows evidence for recent star formation
\citep{derijcke03}. This is reflected in the high PaT absorption
strength, and the corrected CaT* agrees very well with the model
prediction based on FCC\,046's metallicity. the values of the other
Fornax dE with emission, FCC\,207, agree perfectly well with the model
predictions. Surprisingly, in DW\,2 we also found emission lines.
However, the CaT of DW\,2 is already too low compared to the models,
and a high PaT would aggravate the situation for CaT*.  Finally,
FCC\,245 appears to be quite normal, morphologically and
photometrically speaking, but its CaT is also lower than expected for
its age and metallicity.  Still, 13 out of 15 of the observed dE's now
have measured ages, metallicities, and CaT* values that agree with
model predictions. The fact that CaT* values predicted using
age/metallicity estimates that were derived from spectral features in
the blue part of the spectrum ($\sim$4700 -- 5500\,\AA) agree quite
reasonably with the observed CaT* values in the NIR ($\sim$8500\,\AA)
indicates that the SSP assumption is not the cause of the CaT puzzle
for dE's.

\section{Discussion and conclusions} \label{disc}

At least in the dE regime, the \ionCaT\ Triplet puzzle seems to be
solved. With the new age and metallicity estimates presented in this
Letter, the predicted and observed CaT* indices are in good agreement
for all sample galaxies but two. The fact that CaT* values predicted
using age/metallicity estimates that were derived from spectral
features in the blue part of the spectrum agree with the observed CaT*
values in the NIR indicates that the SSP assumption is not the cause
of the CaT puzzle for dE's. Rather, the CaT puzzle in the dE regime
was caused by the spuriously low metallicities and high ages, derived
from lower resolution spectra using less sophisticated theoretical
models, assigned to Fornax dE's. This shows that the CaT* index, and,
in old stellar systems in which the PaT index is small, the CaT index
as well, is indeed a good tracer of metallicity.

This also solves the apparent dichotomy between dE's on the one hand
and globular clusters, Local Group dwarf spheroidals (dSph), and
ultra-compact dwarfs (UCD) on the other hand. CaT line-strengths
measured in individual stars of Local Group dSphs have been shown to
be a very accurate tracer of metallicity \citep{b06,to03} and
[Fe/H]-values derived from CaT measurements have been used extensively
to construct metallicity distributions of the stars in dSphs. Also,
for UCDs \citep{ev07} and globular clusters \citep{saglia02} the CaT
index has proved to be an excellent tracer of metallicity. Here, we
have shown that in dE's as well, the CaT* index measured from 
integrated-light spectra can be used as a tracer of metallicity.


To summarise, we derive new age and metallicity estimates for 16 dE's
in the Fornax Cluster and in Southern Groups using high S/N optical
VLT/FORS1+2 spectra. We have measured the H$\beta$, Mg$b$, Fe5270, and
Fe5335 indices in the Lick/IDS system and applied the TMB03 models to
them. We find that these dE's have solar [$\alpha$/Fe] abundance
ratios. A full-spectrum fit using Pegase-HR with the ELODIE.3.1
stellar library provides us with a second, independent age and
metallicity estimate for these galaxies. We find both approaches to be
in excellent agreement. With mean metallicity ${\rm [Z/H]} =
-0.35$\,dex and ages younger than $\approx 7$\,Gyr, these dE's are
more metal-rich and younger than previously thought.  Some even show
strong emission lines, an indication of on-going star formation, in
agreement with previous H$\alpha$ imaging of dE's
\citep{derijcke03,michielsen04}. The ages and metallicities we derive
for the dE's in the Fornax cluster and in Southern groups fall in
roughly the same range as those derived by \citet{g03} and \citet{v04}
for dE's in the Virgo cluster. This is at variance with previous
estimates for Fornax dEs which yielded lower metallicities and higher
ages \citep{heldmould94, rakos01}, based on lower resolution spectra
and less sophisticated theoretical models. The new age and metallicity
estimates are in good agreement with the observed \ionCaT\ triplet
absorption strengths, solving the Calcium puzzle for low-mass systems.

\acknowledgments Based on observations collected at the European Southern
Observatory, Paranal, Chile (Programs 075.B-0179 and 076.B-0196). DM
acknowledges the EU MAGPOP RTN for financial support.


\begin{thebibliography}{}
        
\bibitem[Battaglia et al.(2006)]{b06} Battaglia, G., et al.\ 2006, \aap, 459,
  423

\bibitem[Cardiel et al.(2003)]{cardieletal03} Cardiel, N., Gorgas, J.,
  S{\'a}nchez-Bl{\'a}zquez, P., Cenarro, A.~J., Pedraz, S., Bruzual,
  G., \& Klement, J.\ 2003, \aap, 409, 511

\bibitem[Cenarro et al.(2004)]{cenarro04} Cenarro, A.~J.,
  S{\'a}nchez-Bl{\'a}zquez, P., Cardiel, N., \& Gorgas, J.\ 2004, \apjl, 614,
  L101

\bibitem[Cenarro et al.(2003)]{cenarro03} Cenarro, A.~J., Gorgas, J.,
  Vazdekis, A., Cardiel, N., \& Peletier, R.~F.\ 2003, \mnras, 339, L12

\bibitem[Cenarro et al.(2001)]{cenarro01} Cenarro, A.~J., Cardiel, N., Gorgas,
  J., Peletier, R.~F., Vazdekis, A., \& Prada, F.\ 2001, \mnras, 326, 959

\bibitem[de Rijcke et al.(2005)]{derijcke05} de Rijcke, S., Michielsen, D.,
  Dejonghe, H., Zeilinger, W.~W., \& Hau, G.~K.~T.\ 2005, \aap, 438, 491

\bibitem[De Rijcke et al.(2003)]{derijcke03} De Rijcke, S., Zeilinger, W.~W.,
  Dejonghe, H., \& Hau, G.~K.~T.\ 2003, \mnras, 339, 225

\bibitem[Evstigneeva et al.(2007)]{ev07} Evstigneeva, E.~A., Gregg, M.~D.,
  Drinkwater, M.~J., \& Hilker, M.\ 2007, \aj, 133, 1722

\bibitem[Falc{\'o}n-Barroso et al.(2003)]{falcon03} Falc{\'o}n-Barroso, J.,
  Peletier, R.~F., Vazdekis, A., \& Balcells, M.\ 2003, \apjl, 588, L17

\bibitem[Geha et al.(2003)]{g03} Geha, M., Guhathakurta, P., \& van der Marel,
  R.~P.\ 2003, \aj, 126, 1794

\bibitem[Held \& Mould(1994)]{heldmould94} Held, E.~V., \& Mould, J.~R.\ 1994,
  \aj, 107, 1307 (HM94)

\bibitem[Koleva et al.(2007)]{koleva07} Koleva, M., Prugniel, P., Ocvirk, P.,
  \& Le Borgne, D.\ 2007, in the proceedings of IAUS241, "Stellar Populations
  as Building Blocks of Galaxies", eds. A. Vazdekis and R. Peletier

\bibitem[Le Borgne et al.(2004)]{leborgne04} Le Borgne, D., Rocca-Volmerange,
  B., Prugniel, P., Lan{\c c}on, A., Fioc, M., \& Soubiran, C.\ 2004, \aap,
  425, 881

\bibitem[Michielsen et al.(2004)]{michielsen04} Michielsen, D., De Rijcke, S.,
  Zeilinger, W. W., Prugniel, P., Dejonghe, H., Roberts, S., 2004, MNRAS, 353,
  1293

\bibitem[Michielsen et al.(2003)]{michielsen03} Michielsen, D., De Rijcke, S.,
  Dejonghe, H., Zeilinger, W.~W., \& Hau, G.~K.~T.\ 2003, \apjl, 597, L21
  (Paper~1)

\bibitem[Pasquali et al.(2005)]{pasquali05} Pasquali, A., Larsen, S.,
  Ferreras, I., Gnedin, O.~Y., Malhotra, S., Rhoads, J.~E., Pirzkal, N., \&
  Walsh, J.~R.\ 2005, \aj, 129, 148

\bibitem[Prochaska et al.(2005)]{prochaskaetal05} Prochaska L.~C., Rose J.~A.,
  \& Schiavon R.~P., 2005, \aj, 130, 2666

\bibitem[Prugniel et al.(2007)]{prugniel07} Prugniel, P., Soubiran, C.,
  Koleva, M., \& Le Borgne, D.\ 2007, ArXiv Astrophysics e-prints,
  arXiv:astro-ph/0703658

\bibitem[Rakos et al.(2001)]{rakos01} Rakos, K., Schombert, J., Maitzen,
  H.~M., Prugovecki, S., \& Odell, A. \ 2001, \aj, 121, 1974 (R01)

\bibitem[Saglia et al.(2002)]{saglia02} Saglia, R.~P., Maraston, C., Thomas,
  D., Bender, R., \& Colless, M.\ 2002, \apjl, 579, L13
  
\bibitem[Tolstoy et al.(2003)]{to03} Tolstoy, E. Irwin, M. J., Cole,
A. A., Pasquini, L., Gilmozzi, R., Gallagher, J. S., 2001, \mnras,
327, 918

\bibitem[Thomas et al.(2003a)]{TMB03} Thomas, D., Maraston, C., \& Bender, R.\ 
  2003a, \mnras, 339, 897 (TMB03)

\bibitem[Thomas et al.(2003b)]{thomas03Ca} Thomas, D., Maraston, C., \&
  Bender, R.\ 2003b, \mnras, 343, 279

\bibitem[van Zee et al.(2004)]{v04} van Zee, L., Barton, E. J., Skillman, E.
  D., 2004, \aj, 128, 2797

\bibitem[Vazdekis et al.(2003)]{vazdekis03} Vazdekis, A., Cenarro, A.~J.,
  Gorgas, J., Cardiel, N., \& Peletier, R.~F.\ 2003, \mnras, 340, 1317

\bibitem[Vazdekis et al.(1997)]{vazdekis97} Vazdekis, A., Peletier, R.~F.,
  Beckman, J.~E., \& Casuso, E.\ 1997, \apjs, 111, 203

\bibitem[\protect\citeauthoryear{Worthey et al.}{1994}]{wortheyetal94} Worthey
  G., Faber S.~M., Gonzalez J.~J., Burstein D., 1994, ApJS, 94, 687

\bibitem[Worthey \& Ottaviani(1997)]{wo97} Worthey, G., \& Ottaviani, D.~L.\ 
  1997, \apjs, 111, 377

\end{thebibliography}
\end{document}